%% file: gpce.tex
\documentclass[sigplan]{acmart}
\usepackage[utf8]{inputenc}
\usepackage{color}
\usepackage{epsfig}

\usepackage{microtype}
\usepackage{balance}
\DeclareUnicodeCharacter{21D2}{$\Rightarrow$}
\DeclareUnicodeCharacter{21DD}{$\rightsquigarrow$}
\DeclareUnicodeCharacter{2026}{\textrm{\ldots}}
\acmDOI{10.1145/3136040.3136045}
\acmYear{2017}
\copyrightyear{2017}
\acmConference[GPCE'17]{16th ACM SIGPLAN International Conference on Generative Programming: Concepts and Experiences}{October 23--24, 2017}{Vancouver, Canada}

\let\oldthing\footnotetextcopyrightpermission
\renewcommand\footnotetextcopyrightpermission[1]{\oldthing{
  Copyright \copyright 2017 Ludovic Courtès.\

  Permission is granted to copy, distribute and/or modify this document
  under the terms of the GNU Free Documentation License, Version 1.3
  or any later version published by the Free Software Foundation;
  with no Invariant Sections, no Front-Cover Texts, and no Back-Cover Texts.
  A copy of the license is
  available at \url{https://www.gnu.org/licenses/gfdl.html}.

  The source of this document is available from
  \url{https://git.sv.gnu.org/cgit/guix/maintenance.git}.
}}

\newdimen\oldframetabcolsep
\newdimen\oldcolortabcolsep
\newdimen\oldpretabcolsep

\title{Code Staging in GNU Guix}
\author{Ludovic Court\`{e}s}
\affiliation{
  \institution{Inria}
  \city{Bordeaux, France}}
\begin{document}
\begin{abstract}
GNU Guix is a “functional” package manager that builds upon %
earlier work on Nix.  Guix implements high-level abstractions such as %
packages and operating system services as domain-specific languages %
(DSLs) embedded in Scheme.  It also implements build actions and %
operating system orchestration in Scheme.  This leads to a multi-tier %
programming environment where embedded code snippets are staged for %
eventual execution.\par
This paper presents {\em{G-expressions}} or “{\em{gexps}}”, the staging mechanism we devised for Guix.  We explain our %
journey from traditional Lisp S-expressions to G-expressions, which %
augment the former with contextual information and ensure hygienic code staging. %
We discuss the %
implementation of gexps and report on our experience using them in a %
variety of operating system use cases—from package build processes %
to system services.  Gexps %
provide a novel way to cover many aspects of OS configuration in a %
single, multi-tier language, while facilitating code reuse and code %
sharing.\par

\end{abstract}
\input{categories.tex}
\maketitle

\section{Introduction}
\label{chapter4286}
Users of free operating systems such as GNU/Linux are used %
to {\em{package managers}} like Debian's {\texttt{apt-\-get}}, which %
allow them to install, upgrade, and remove software from a large %
collection of free software packages.  GNU Guix\footnote{{\texttt{https{\char58}/\-/\-gnu.\-org/\-software/\-guix}}} is a {\em{functional}} package %
manager that builds upon the ideas developed for Nix by Dolstra {\textit{et al.}} [5].  The term “functional” means %
that software build processes are considered as pure functions{\char58} given %
a set of inputs (compiler, libraries, build scripts, and so on), a %
package’s build function is assumed to always produce the same result. %
Build results are stored in an immutable persistent data structure, %
the {\em{store}}, implemented as a single directory, {\texttt{/\-gnu/\-store}}. %
Each entry in {\texttt{/\-gnu/\-store}} has a file name composed of the hash %
of all the build inputs used to produce it, followed by a symbolic %
name.  For example, {\texttt{/\-gnu/\-store/\-yr9rk90jf…-\-gcc-\-7.\-1.\-0}} identifies %
a specific build of GCC 7.1.  A variant of GCC 7.1, for instance one %
using different build options or different dependencies, would get a %
different hash.  Thus, each store file name uniquely identifies build %
results, and build processes are {\em{referentially transparent}}. %
This simplifies reasoning on complex package compositions, and %
also has nice properties such as supporting transactional upgrades and %
rollback “for free.”  The %
Guix System Distribution (or GuixSD) and NixOS extend the %
functional paradigm to whole operating system deployments [6].\par
Guix implements this functional deployment paradigm %
pioneered by Nix but, as explained in previous work, its %
implementation departs from Nix in interesting ways [3].  First, while Nix relies on a custom %
domain-specific language (DSL), the Nix language, Guix instead %
implements a set of DSLs and data structures embedded in the %
general-purpose language Scheme.  This simplifies %
the development of user interfaces and tools, %
and allows users to benefit from everything a %
general-purpose language brings{\char58} compiler, debugger, REPL, editor %
support, libraries, and so on.\par
\begin{figure}[ht]
\begin{scriptsize}
\setlength{\oldpretabcolsep}{\tabcolsep}
\addtolength{\tabcolsep}{-\tabcolsep}
{\setbox1 \vbox \bgroup
{\noindent \texttt{\begin{tabular}{l}
({\textcolor[rgb]{0.4117647058823529,0.34901960784313724,0.8117647058823529}{{\textbf{define}}}}\ hello\\
\ \ (package\\
\ \ \ \ (name\ {\textcolor[rgb]{1.0,0.0,0.0}{"hello"}})\\
\ \ \ \ (version\ {\textcolor[rgb]{1.0,0.0,0.0}{"2.8"}})\\
\ \ \ \ (source\ (origin\ \ \ \ \ \ \ \ \ \ \ \ \ \ \ \ \ \ \ \ \ \ \ \ \ \ \ \ {\textcolor[rgb]{0.4666666666666667,0.4,0.0}{{\textit{\textrm{}}}}}\\
\ \ \ \ \ \ \ \ \ \ \ \ \ \ (method\ url-fetch)\\
\ \ \ \ \ \ \ \ \ \ \ \ \ \ (uri\ (string-append\ {\textcolor[rgb]{1.0,0.0,0.0}{"mirror{\char58}//gnu/hello/hello-"}}\\
\ \ \ \ \ \ \ \ \ \ \ \ \ \ \ \ \ \ \ \ \ \ \ \ \ \ \ \ \ \ \ \ \ \ version\ {\textcolor[rgb]{1.0,0.0,0.0}{".tar.gz"}}))\ {\textcolor[rgb]{0.4666666666666667,0.4,0.0}{{\textit{\textrm{}}}}}\\
\ \ \ \ \ \ \ \ \ \ \ \ \ \ (sha256\ (base32\ {\textcolor[rgb]{1.0,0.0,0.0}{"0wqd8…"}}))))\ \ \ \ \ {\textcolor[rgb]{0.4666666666666667,0.4,0.0}{{\textit{\textrm{}}}}}\\
\ \ \ \ (build-system\ gnu-build-system)\ \ \ \ \ \ \ \ \ \ \ \ {\textcolor[rgb]{0.4666666666666667,0.4,0.0}{{\textit{\textrm{}}}}}\\
\ \ \ \ (arguments\\
\ \ \ \ \ \ '(\#{\char58}configure-flags\ \ \ \ \ \ \ \ \ \ \ \ \ \ \ \ \ \ \ \ \ \ {\textcolor[rgb]{0.4666666666666667,0.4,0.0}{{\textit{\textrm{}}}}}\\
\ \ \ \ \ \ \ \ `({\textcolor[rgb]{1.0,0.0,0.0}{"--disable-color"}}\\
\ \ \ \ \ \ \ \ \ \ ,(string-append\ {\textcolor[rgb]{1.0,0.0,0.0}{"--with-gawk="}}\\
\ \ \ \ \ \ \ \ \ \ \ \ \ \ \ \ \ \ \ \ \ \ \ \ \ \ (assoc-ref\ \%build-inputs\ {\textcolor[rgb]{1.0,0.0,0.0}{"gawk"}})))))\\
\ \ \ \ (inputs\ `(({\textcolor[rgb]{1.0,0.0,0.0}{"gawk"}}\ ,gawk)))\ \ \ \ \ \ \ \ \ \ \ \ \ \ \ \ {\textcolor[rgb]{0.4666666666666667,0.4,0.0}{{\textit{\textrm{}}}}}\\
\ \ \ \ (synopsis\ {\textcolor[rgb]{1.0,0.0,0.0}{"GNU\ Hello"}})\\
\ \ \ \ (description\ {\textcolor[rgb]{1.0,0.0,0.0}{"Example\ of\ a\ GNU\ package."}})\\
\ \ \ \ (home-page\ {\textcolor[rgb]{1.0,0.0,0.0}{"https{\char58}//gnu.org/software/hello/"}})\\
\ \ \ \ (license\ gpl3+)))\\
\end{tabular}
}}
\egroup{\box1}}%
\setlength{\tabcolsep}{\oldpretabcolsep}

\end{scriptsize}
\caption{\label{fig-package-def}A package definition using the high-level %
interface.}\end{figure}
In Guix, high-level package definitions like the one shown %
in Figure \ref{fig-package-def} are compiled to {\em{derivations}}, the low-level representation of build actions %
inherited from Nix.  A derivation specifies{\char58} a command to run to %
perform the build (the {\em{build program}}), environment variables to be %
defined, and derivations whose build result it depends on. %
Derivations are sent to a privileged daemon, which is responsible for %
building them on behalf of clients.  The build daemon creates isolated %
environments ({\em{containers}} in a chroot) in which it spawns %
the build program; isolated build environments ensure %
that build programs do not depend on undeclared inputs.\par
The second way in which Guix departs from Nix is by using %
the same language, Scheme, for all its functionality.  While package %
definitions in Nix can embed Bash or Perl snippets to refine build %
steps, Guix package definitions instead embed Scheme code. %
Consequently, we have two strata of Scheme code{\char58} the {\em{host code}}, %
which provides the package definition, and the {\em{build code}}, which is %
staged for later execution by the build daemon.\par
This paper focuses on code staging in Guix.  Our contribution %
is twofold{\char58} we present G-expressions (or “gexps”), a new code staging %
mechanism implemented through mere syntactic extensions of the Scheme %
language; we show the use of gexps in several areas of the “orchestration” programs %
of the operating system.  Section~2 %
discusses the early attempt at code staging in Guix, as %
mentioned in [3], and its shortcomings. %
Section~3 presents the design and %
implementation of gexps. %
Section~4 reports on our experience using %
gexps in a variety of areas in Guix and GuixSD.  Section~5 discusses limitations and future work. %
Finally Section~6 compares gexps to %
related work and Section~7 %
concludes.\par

\section{Early Attempt}
\label{origins}
Scheme is a dialect of Lisp, and Lisp is famous for its %
its direct representation of code as a data %
structure using the same syntax.  “S-expressions” or “sexps”, Lisp’s %
parenthecal expressions, thus look like they lend themselves to code %
staging. %
In this section we show how our early experience made it clear that %
we needed an {\em{augmented}} version of sexps.\par

\subsection{Staging Build Expressions}
\label{section4377}
\begin{figure}[ht]
\begin{scriptsize}
\setlength{\oldpretabcolsep}{\tabcolsep}
\addtolength{\tabcolsep}{-\tabcolsep}
{\setbox1 \vbox \bgroup
{\noindent \texttt{\begin{tabular}{l}
(\textbf{\textrm{let*}}\ ((store\ (open-connection))\\
\ \ \ \ \ \ \ (drv\ \ \ (package-derivation\ store\ imagemagick))\\
\ \ \ \ \ \ \ (image\ (add-to-store\ store\ {\textcolor[rgb]{1.0,0.0,0.0}{"image.png"}}\ \#t\ {\textcolor[rgb]{1.0,0.0,0.0}{"sha256"}}\\
\ \ \ \ \ \ \ \ \ \ \ \ \ \ \ \ \ \ \ \ \ \ \ \ \ \ \ \ {\textcolor[rgb]{1.0,0.0,0.0}{"./GuixSD.png"}}))\\
\ \ \ \ \ \ \ (build\\
\ \ \ \ \ \ \ \ '(\textbf{\textrm{let}}\ ((imagemagick\ (assoc-ref\ \%build-inputs\\
\ \ \ \ \ \ \ \ \ \ \ \ \ \ \ \ \ \ \ \ \ \ \ \ \ \ \ \ \ \ \ \ \ \ \ \ \ \ \ {\textcolor[rgb]{1.0,0.0,0.0}{"imagemagick"}}))\\
\ \ \ \ \ \ \ \ \ \ \ \ \ \ \ (image\ (assoc-ref\ \%build-inputs\ {\textcolor[rgb]{1.0,0.0,0.0}{"image"}})))\\
\ \ \ \ \ \ \ \ \ \ \ (mkdir\ \%output)\\
\ \ \ \ \ \ \ \ \ \ \ (system*\ (string-append\ imagemagick\ {\textcolor[rgb]{1.0,0.0,0.0}{"/bin/convert"}})\\
\ \ \ \ \ \ \ \ \ \ \ \ \ \ \ \ \ \ \ \ {\textcolor[rgb]{1.0,0.0,0.0}{"-quality"}}\ {\textcolor[rgb]{1.0,0.0,0.0}{"75\%"}}\\
\ \ \ \ \ \ \ \ \ \ \ \ \ \ \ \ \ \ \ \ image\\
\ \ \ \ \ \ \ \ \ \ \ \ \ \ \ \ \ \ \ \ (string-append\ \%output\ {\textcolor[rgb]{1.0,0.0,0.0}{"/image.jpg"}})))))\\
\ \ (build-expression-$>$derivation\ store\ {\textcolor[rgb]{1.0,0.0,0.0}{"example"}}\ build\\
\ \ \ \ \ \ \ \ \ \ \ \ \ \ \ \ \ \ \ \ \ \ \ \ \ \ \ \ \ \ \ \ \#{\char58}inputs\ `(({\textcolor[rgb]{1.0,0.0,0.0}{"imagemagick"}}\ ,drv)\\
\ \ \ \ \ \ \ \ \ \ \ \ \ \ \ \ \ \ \ \ \ \ \ \ \ \ \ \ \ \ \ \ \ \ \ \ \ \ \ \ \ \ \ ({\textcolor[rgb]{1.0,0.0,0.0}{"image"}}\ ,image))))\\
\end{tabular}
}}
\egroup{\box1}}%
\setlength{\tabcolsep}{\oldpretabcolsep}

\end{scriptsize}
\caption{\label{fig-build-sexp}First attempt{\char58} build expressions as sexps.}\end{figure}
In previous work [3], we %
presented our first attempt at writing build expressions in Scheme, %
which relied solely on Lisp quotation [2].  Figure \ref{fig-build-sexp} %
shows an example that creates a derivation that, when built, converts %
the input image to JPEG, using the {\texttt{convert}} program from the %
ImageMagick package—this is equivalent to a three-line makefile rule, but %
referentially transparent.  In this example, variable {\texttt{store}} %
represents the connection to the build daemon.  The {\texttt{package-\-derivation}} function takes the {\texttt{imagemagick}} package %
object and computes its corresponding derivation, while the {\texttt{add-\-to-\-store}} remote procedure call (RPC) instructs the daemon to %
add the file {\texttt{GuixSD.\-png}} to {\texttt{/\-gnu/\-store}}.  The variable %
{\texttt{build}} contains our build program as an sexp (the %
apostrophe is equivalent to {\texttt{quote}}; it introduces unevaluated %
code).  Finally, {\texttt{build-\-expression-\-$>$derivation}} takes the build program and computes %
the corresponding derivation without building it.  The user can then %
make an RPC to the daemon asking it to build this derivation; %
only then will the daemon create an isolated environment and run our %
build program.\par
{\texttt{build-\-expression-\-$>$derivation}} arranges so that the %
build program, {\texttt{build}}, is evaluated in a context where two %
extra variables are defined{\char58} {\texttt{\%build-\-inputs}} contains a list %
that maps labels, such as {\texttt{"imagemagick"}}, to file names, such %
as {\texttt{/\-gnu/\-store/\-…-\-imagemagick-\-6.\-9}}, and {\texttt{\%output}} contains %
the file name for the result of this derivation.  Thus, the {\texttt{assoc-\-ref}} calls in {\texttt{build}} allow it to retrieve the file name %
of ImageMagick.\par

\subsection{S-Expressions Are Not Enough}
\label{section4392}
Needless to say, this interface was extremely verbose and %
inconvenient—fortunately, users usually only had to deal with the more %
pleasant {\texttt{package}} interface shown in Figure \ref{fig-package-def}.  All these steps are necessary to define the %
derivation and its dependencies, but where does the verbosity come %
from?  First, we have to explicitly call {\texttt{package-\-derivation}} %
for each package the expression refers to.  Second, we have to %
specify the inputs with labels at the call site.  Third, the build %
code has to use this {\texttt{assoc-\-ref}} call just to retrieve the {\texttt{/\-gnu/\-store}} file name of its inputs.  It is error-prone{\char58} if we %
omit the {\texttt{\#{\char58}inputs}} parameter, of if we mispell an input label, %
we will only find out when we build the derivation.\par
Another limitation not visible on a toy example but that %
became clear as we developed GuixSD is the cost of carrying this {\texttt{\#{\char58}inputs}} argument down to the call site.  It forces programmers to %
carry not only the build expression, {\texttt{build}}, but also the %
corresponding {\texttt{inputs}} argument, and %
makes it very hard to compose build expressions.\par
While {\texttt{quote}} allowed us to easily represent code, it %
clearly lacked some of the machinery that would make %
staging in Guix more convenient.  It boils down to two things{\char58} it %
lacks {\em{context}}—the set of inputs associated with the %
expression—and it lacks the ability to serialize high-level objects—to %
replace a reference to a package object with its {\texttt{/\-gnu/\-store}} %
file name.\par

\section{G-Expressions}
\label{gexps}
We devised “G-expressions” to address %
these shortcomings.  This section describes the design and implementation of %
G-expressions, as well as extensions we added to address new use %
cases.\par

\subsection{Design Principle}
\label{section4486}
\begin{figure}[ht]
\begin{scriptsize}
\setlength{\oldpretabcolsep}{\tabcolsep}
\addtolength{\tabcolsep}{-\tabcolsep}
{\setbox1 \vbox \bgroup
{\noindent \texttt{\begin{tabular}{l}
(\textbf{\textrm{let*}}\ ((image\ (local-file\ {\textcolor[rgb]{1.0,0.0,0.0}{"./GuixSD.png"}}\ {\textcolor[rgb]{1.0,0.0,0.0}{"image.png"}}))\\
\ \ \ \ \ \ \ (build\ \#$_{\mbox{\char126}}$(\textbf{\textrm{begin}}\\
\ \ \ \ \ \ \ \ \ \ \ \ \ \ \ \ \ \ (mkdir\ \#\$output)\\
\ \ \ \ \ \ \ \ \ \ \ \ \ \ \ \ \ \ (system*\ (string-append\ \#\$imagemagick\\
\ \ \ \ \ \ \ \ \ \ \ \ \ \ \ \ \ \ \ \ \ \ \ \ \ \ \ \ \ \ \ \ \ \ \ \ \ \ \ \ \ \ {\textcolor[rgb]{1.0,0.0,0.0}{"/bin/convert"}})\\
\ \ \ \ \ \ \ \ \ \ \ \ \ \ \ \ \ \ \ \ \ \ \ \ \ \ \ {\textcolor[rgb]{1.0,0.0,0.0}{"-quality"}}\ {\textcolor[rgb]{1.0,0.0,0.0}{"75\%"}}\\
\ \ \ \ \ \ \ \ \ \ \ \ \ \ \ \ \ \ \ \ \ \ \ \ \ \ \ \#\$image\\
\ \ \ \ \ \ \ \ \ \ \ \ \ \ \ \ \ \ \ \ \ \ \ \ \ \ \ (string-append\ \#\$output\ {\textcolor[rgb]{1.0,0.0,0.0}{"/image.jpg"}})))))\\
\ \ (gexp-$>$derivation\ {\textcolor[rgb]{1.0,0.0,0.0}{"example"}}\ build))\\
\end{tabular}
}}
\egroup{\box1}}%
\setlength{\tabcolsep}{\oldpretabcolsep}

\end{scriptsize}
\caption{\label{fig-build-gexp}Creating a derivation from a gexp.}\end{figure}
G-expressions {\em{bind software deployment to %
staging}}{\char58} when a gexp is staged, the software and artifacts it refers %
to are guaranteed to be deployed as well. %
A gexp bundles an sexp and its inputs %
and outputs, and it can be serialized with {\texttt{/\-gnu/\-store}} file %
names substituted as needed.  We first define two operators{\char58} %
\begin{itemize}
 \item {\texttt{gexp}}, abbreviated {\texttt{\#$_{\mbox{\char126}}$}}, is the counterpart of %
Scheme’s {\texttt{quasiquote}}{\char58} it allows users to describe unevaluated %
code.
 \item {\texttt{ungexp}}, abbreviated {\texttt{\#\$}}, is the counterpart %
of Scheme’s {\texttt{unquote}}{\char58} it allows quoted code to refer to values in %
the host program.  These values can be of any of Scheme’s primitive %
data types, but we are specifically interested in values such as %
package objects that can be “compiled” to elements in the store.
 \item {\texttt{ungexp-\-splicing}}, abbreviated {\texttt{\#\$@}}, allows a %
list of elements to be “spliced” in the surrounding list, similar to %
Scheme’s {\texttt{unquote-\-splicing}}.
\end{itemize}  %
The example in Figure \ref{fig-build-sexp}, rewritten as a %
gexp, is shown in Figure \ref{fig-build-gexp}.  We have all %
the properties we were looking for{\char58} the gexp carries %
information about its inputs that does not need to be passed at the %
{\texttt{gexp-\-$>$derivation}} call site, and the reference to {\texttt{imagemagick}}, which is bound to a package object, is automatically %
rewritten to the corresponding store file name by {\texttt{gexp-\-$>$derivation}}\footnote{Compared to Figure \ref{fig-build-sexp}, the {\texttt{store}} argument has disappeared.  This %
is because we implemented {\texttt{gexp-\-$>$derivation}} as a monadic %
function in the {\em{state monad}}, where the state threaded %
through monadic function calls is that store parameter.  The use of a %
monadic interface is completely orthogonal to the gexp design though, %
so we will not insist on it.}.  {\texttt{local-\-file}} returns a new %
record that denotes a file from the local file system to be %
added to the store.\par
Under the hood, {\texttt{gexp-\-$>$derivation}} converts the %
gexp to an sexp, the residual build program, and stores it under {\texttt{/\-gnu/\-store}}.  In doing that, it replaces the {\texttt{ungexp}} forms %
{\texttt{\#\$imagemagick}} and {\texttt{\#\$image}} with their corresponding %
{\texttt{/\-gnu/\-store}} file names.  The special {\texttt{\#\$output}} form, %
which is used to refer to the output file name of the derivation, is %
itself replaced by a {\texttt{getenv}} call to retrieve its value at run %
time (a necessity since the output file name is not known at the time %
the gexp is serialized to disk.)\par
The {\texttt{gexp-\-$>$derivation}} function has an optional %
{\texttt{\#{\char58}system}} argument that can be used to specify the system on %
which the derivation is to be built.  For instance, passing {\texttt{\#{\char58}system "i686-\-linux"}} forces a 32-bit build on a GNU system running %
the kernel Linux.  The gexp itself is system-independent; it refers to %
the {\texttt{imagemagick}} package object, which is also %
system-independent.  Since the store file name of derivation outputs %
is a function of the system type, {\texttt{gexp-\-$>$derivation}} must make %
sure to use the file name of ImageMagick corresponding to its {\texttt{\#{\char58}system}} argument.  Therefore, this substitution must happen when %
{\texttt{gexp-\-$>$derivation}} is invoked, and {\em{not}} when the gexp %
is created.\par
G-expressions are “hygienic”{\char58} {\em{they preserve %
lexical scope across stages}} [9, 13, 15].\begin{figure}[ht]
\begin{scriptsize}
\setlength{\oldpretabcolsep}{\tabcolsep}
\addtolength{\tabcolsep}{-\tabcolsep}
{\setbox1 \vbox \bgroup
{\noindent \texttt{\begin{tabular}{l}
(\textbf{\textrm{let}}\ ((gen-body\ (\textbf{\textrm{lambda}}\ (x)\\
\ \ \ \ \ \ \ \ \ \ \ \ \ \ \ \ \ \ \#$_{\mbox{\char126}}$(\textbf{\textrm{let}}\ ((x\ 40))\\
\ \ \ \ \ \ \ \ \ \ \ \ \ \ \ \ \ \ \ \ \ \ (+\ x\ \#\$x)))))\\
\ \ \#$_{\mbox{\char126}}$(\textbf{\textrm{let}}\ ((x\ 2))\\
\ \ \ \ \ \ \#\$(gen-body\ \#$_{\mbox{\char126}}$x))\\
\\
⇝\ (\textbf{\textrm{let}}\ ((x-1bd8-0\ 2))\\
\ \ \ \ (\textbf{\textrm{let}}\ ((x-4f05-0\ 40))\ (+\ x-4f05-0\ x-1bd8-0)))\\
\end{tabular}
}}
\egroup{\box1}}%
\setlength{\tabcolsep}{\oldpretabcolsep}

\end{scriptsize}
\caption{\label{fig-gexp-hygiene}Lexical scope preservation across stages (⇝ %
denotes code generation).}\end{figure}
Figure \ref{fig-gexp-hygiene} illustrates two %
well-known properties of hygienic multi-stage programs{\char58} first, binding %
{\texttt{x}} in one stage (outside the gexp) is distinguished from %
binding {\texttt{x}} in another stage (inside the gexp); second, binding %
{\texttt{x}} introduced inside {\texttt{gen-\-body}} does not shadow binding %
{\texttt{x}} in the outer gexp thanks to the renaming of these variables %
in the residual program.\par

\subsection{Implementation}
\label{implementation}
\begin{figure}[ht]
\begin{scriptsize}
\setlength{\oldpretabcolsep}{\tabcolsep}
\addtolength{\tabcolsep}{-\tabcolsep}
{\setbox1 \vbox \bgroup
{\noindent \texttt{\begin{tabular}{l}
\#$_{\mbox{\char126}}$(list\ (string-append\ \#\$imagemagick\ {\textcolor[rgb]{1.0,0.0,0.0}{"/bin/convert"}})\\
\ \ \ \ \ \ \ \ (string-append\ \#\$emacs\ {\textcolor[rgb]{1.0,0.0,0.0}{"/bin/emacs"}}))\\
\\
\\
⇒\ (\textbf{\textrm{gexp}}\ (list\ (string-append\ (\textbf{\textrm{ungexp}}\ imagemagick)\\
\ \ \ \ \ \ \ \ \ \ \ \ \ \ \ \ \ \ \ \ \ \ \ \ \ \ \ \ \ {\textcolor[rgb]{1.0,0.0,0.0}{"/bin/convert"}})\\
\ \ \ \ \ \ \ \ \ \ \ \ \ \ (string-append\ (\textbf{\textrm{ungexp}}\ emacs)\\
\ \ \ \ \ \ \ \ \ \ \ \ \ \ \ \ \ \ \ \ \ \ \ \ \ \ \ \ \ {\textcolor[rgb]{1.0,0.0,0.0}{"/bin/emacs"}})))\\
\\
⇒\ (\textbf{\textrm{let}}\ ((references\\
\ \ \ \ \ \ \ \ \ (list\ (gexp-input\ imagemagick)\\
\ \ \ \ \ \ \ \ \ \ \ \ \ \ \ (gexp-input\ emacs)))\\
\ \ \ \ \ \ \ \ (proc\ (\textbf{\textrm{lambda}}\ (a\ b)\\
\ \ \ \ \ \ \ \ \ \ \ \ \ \ \ \ (list\ 'list\\
\ \ \ \ \ \ \ \ \ \ \ \ \ \ \ \ \ \ \ \ \ \ (list\ 'string-append\ a\\
\ \ \ \ \ \ \ \ \ \ \ \ \ \ \ \ \ \ \ \ \ \ \ \ \ \ \ \ {\textcolor[rgb]{1.0,0.0,0.0}{"/bin/convert"}})\\
\ \ \ \ \ \ \ \ \ \ \ \ \ \ \ \ \ \ \ \ \ \ (list\ 'string-append\ b\\
\ \ \ \ \ \ \ \ \ \ \ \ \ \ \ \ \ \ \ \ \ \ \ \ \ \ \ \ {\textcolor[rgb]{1.0,0.0,0.0}{"/bin/emacs"}})))))\\
\ \ \ \ (make-gexp\ references\ proc))\\
\\
⇝\ (list\ (string-append\ {\textcolor[rgb]{1.0,0.0,0.0}{"/gnu/store/65qrc…-imagemagick-6.9"}}\\
\ \ \ \ \ \ \ \ \ \ \ \ \ \ \ \ \ \ \ \ \ \ \ {\textcolor[rgb]{1.0,0.0,0.0}{"/bin/convert"}})\\
\ \ \ \ \ \ \ \ (string-append\ {\textcolor[rgb]{1.0,0.0,0.0}{"/gnu/store/825n3…-emacs-25.2"}}\\
\ \ \ \ \ \ \ \ \ \ \ \ \ \ \ \ \ \ \ \ \ \ \ {\textcolor[rgb]{1.0,0.0,0.0}{"/bin/emacs"}}))\\
\end{tabular}
}}
\egroup{\box1}}%
\setlength{\tabcolsep}{\oldpretabcolsep}

\end{scriptsize}
\caption{\label{fig-gexp-expansion}Macro expansion (⇒) of a G-expression and code %
generation (⇝).}\end{figure}
As can be seen from the examples above, gexps are %
first-class Scheme values{\char58} a variable can be bound to a gexp, and %
gexps can be passed around like any other value.  The implementation %
consists of two parts{\char58} a syntactic layer that turns {\texttt{\#$_{\mbox{\char126}}$}} forms %
into code that instantiates gexp records, and run-time support %
functions to serialize gexps and to {\em{lower}} their inputs.\par
Scheme is extensible through macros, and {\texttt{gexp}} is a %
{\texttt{syntax-\-case}} macro [7]; {\texttt{\#$_{\mbox{\char126}}$}} and {\texttt{\#\$}} are {\textit{reader %
macros}} that expand to {\texttt{gexp}} or {\texttt{ungexp}} sexps.  This %
is implemented as a library for GNU~Guile, an R5RS/R6RS Scheme %
implementation, {\em{without any modification to its compiler}}. %
Figure \ref{fig-gexp-expansion} shows what our {\texttt{gexp}} %
macro expands to.  In the expanded code, {\texttt{gexp-\-input}} returns a %
record representing a dependency, while {\texttt{make-\-gexp}} returns a %
record representing the whole gexp.  The expanded code defines a %
function of two arguments, {\texttt{proc}}, that returns an sexp; the %
sexp is the body of the gexp with these two arguments inserted %
at the point where the original {\texttt{ungexp}} forms appeared. %
Internally, {\texttt{gexp-\-$>$sexp}}, the function that converts gexps to %
sexps, calls this two-argument procedure passing it the store file %
names of ImageMagick and Emacs.  This strategy gives us constant-time %
substitutions.\par
The internal {\texttt{gexp-\-inputs}} function returns, for a %
given gexp, store, and system type, the derivations that the gexp %
depends on.  In this example, it returns the derivations for %
ImageMagick and Emacs, as computed by the {\texttt{package-\-derivation}} %
function seen earlier.  Gexps can be nested, as in {\texttt{\#$_{\mbox{\char126}}$\#\$\#$_{\mbox{\char126}}$(string-\-append \#\$emacs "/\-bin/\-emacs")}}.  The input list %
returned by {\texttt{gexp-\-inputs}} for the outermost gexp is the sum of %
the inputs of the outermost gexp and the inputs nested gexps.  Likewise, %
{\texttt{gexp-\-outputs}} returns the outputs declared in a gexp and in %
nested gexps.\par
The {\texttt{gexp}} macro performs several passes on its body{\char58} %
\begin{enumerate}
 \item The first pass {\em{\begin{math}\alpha\end{math}-renames lexical %
bindings}} introduced by the gexp in order to preserve lexical scope, %
as illustrated by Figure \ref{fig-gexp-hygiene}.  The %
implementation is similar to MetaScheme [15] and to that described by Rhiger [13], with caveats discussed in Section~5.
 \item The second pass {\em{collects the escape forms}} ({\texttt{ungexp}} variants) in the input source.  The list of escape forms is %
needed to construct the list of inputs stored in the gexp %
record, and to construct the formal argument list of the gexp’s code %
generation function shown in Figure \ref{fig-gexp-expansion}.
 \item The third pass {\em{substitutes escape forms}} with %
references to the corresponding formal arguments of the code %
generation function.  This leads to the sexp-construction expression %
shown in Figure \ref{fig-gexp-expansion}.
\end{enumerate}
Unlike the examples usually given in %
the literature, our renaming pass must generate identifiers in a {\em{deterministic}} fashion{\char58} if they were not, we would produce different %
derivations at each run, which in turn would trigger full rebuilds of %
the package graph.  Thus, instead of relying on {\texttt{gensym}} and %
{\texttt{generate-\-temporaries}}, we generate identifiers as a function of %
the hash of %
the input expression and of the lexical nesting level of %
the identifier—these are the two components we can see in the generated %
identifiers of Figure \ref{fig-gexp-hygiene}.\par
\begin{figure}[ht]
\begin{scriptsize}
\setlength{\oldpretabcolsep}{\tabcolsep}
\addtolength{\tabcolsep}{-\tabcolsep}
{\setbox1 \vbox \bgroup
{\noindent \texttt{\begin{tabular}{l}
({\textcolor[rgb]{0.4117647058823529,0.34901960784313724,0.8117647058823529}{{\textbf{define-gexp-compiler}}}}\ (package-compiler\ (package\ $<$package$>$)\\
\ \ \ \ \ \ \ \ \ \ \ \ \ \ \ \ \ \ \ \ \ \ \ \ \ \ \ \ \ \ \ \ \ \ \ \ \ \ \ \ system\ target)\\
\ \ {\textcolor[rgb]{0.4666666666666667,0.4,0.0}{{\textit{\textrm{;;\ Compile\ PACKAGE\ to\ a\ derivation\ for\ SYSTEM,\ optionally}}}}}\\
\ \ {\textcolor[rgb]{0.4666666666666667,0.4,0.0}{{\textit{\textrm{;;\ cross-compiled\ for\ TARGET.}}}}}\\
\ \ (\textbf{\textrm{if}}\ target\\
\ \ \ \ \ \ (package-$>$cross-derivation\ package\ target\ system)\\
\ \ \ \ \ \ (package-$>$derivation\ package\ system)))\\
\\
({\textcolor[rgb]{0.4117647058823529,0.34901960784313724,0.8117647058823529}{{\textbf{define-gexp-compiler}}}}\ (local-file-compiler\ (file\ $<$local-file$>$)\\
\ \ \ \ \ \ \ \ \ \ \ \ \ \ \ \ \ \ \ \ \ \ \ \ \ \ \ \ \ \ \ \ \ \ \ \ \ \ \ \ \ \ \ system\ target)\\
\ \ {\textcolor[rgb]{0.4666666666666667,0.4,0.0}{{\textit{\textrm{;;\ "Compile"\ FILE\ by\ adding\ it\ to\ the\ store.}}}}}\\
\ \ (\textbf{\textrm{match}}\ file\\
\ \ \ \ ((\$\ $<$local-file$>$\ file\ name)\\
\ \ \ \ \ (interned-file\ file\ name))))\\
\end{tabular}
}}
\egroup{\box1}}%
\setlength{\tabcolsep}{\oldpretabcolsep}

\end{scriptsize}
\caption{\label{fig-gexp-compilers}The gexp compilers for package objects and for %
{\texttt{local-\-file}} objects.}\end{figure}
We have seen that gexps often refer to package objects, %
but we want users to be able to refer to other types of high-level %
objects, such as the {\texttt{local-\-file}} object that appears in Figure %
\ref{fig-gexp-expansion}.  Therefore, the gexp mechanism %
can be extended by defining {\em{gexp compilers}}.  Gexp compilers %
define, for a given data type, how objects of that type can be lowered %
to a derivation or to a literal file in the store.  Figure \ref{fig-gexp-compilers} shows the compilers for objects of the %
package and local-file types.  All these compilers do is call the %
(monadic) function that computes the corresponding derivation, in the %
case of packages, or that simply adds the file to the store in the %
case of {\texttt{local-\-file}}.  Internally, these compilers are invoked %
by {\texttt{gexp-\-$>$sexp}} when it encounters instances of the relevant %
type in a gexp that is being processed.\par
Gexp compilers can also have an associated {\em{expander}}, which specifies how objects should be “rendered” in the %
residual sexp.  The default expander simply produces the store file name %
of the derivation output.  For example, %
assuming the variable {\texttt{emacs}} is bound to a package object, {\texttt{\#$_{\mbox{\char126}}$(string-\-append \#\$emacs "/\-bin/\-emacs")}} expands to {\texttt{(string-\-append "/\-gnu/\-store/\-…-\-emacs-\-25.\-2" "/\-bin/\-emacs")}}, as we have %
seen earlier.  We defined a {\texttt{file-\-append}} function that returns %
objects with a custom expander{\char58} one that performs string concatenation %
when generating the sexp.  We can now write gexps like{\char58}\\[0.3cm]
\begin{scriptsize}
\setlength{\oldpretabcolsep}{\tabcolsep}
\addtolength{\tabcolsep}{-\tabcolsep}
{\setbox1 \vbox \bgroup
{\noindent \texttt{\begin{tabular}{l}
\#$_{\mbox{\char126}}$(execl\ \#\$(file-append\ emacs\ {\textcolor[rgb]{1.0,0.0,0.0}{"/bin/emacs"}}))\\
\\
⇝\ (execl\ {\textcolor[rgb]{1.0,0.0,0.0}{"/gnu/store/…-emacs-25.2/bin/emacs"}})\\
\end{tabular}
}}
\egroup{\box1}}%
\setlength{\tabcolsep}{\oldpretabcolsep}
\end{scriptsize}
\\[0.3cm]
This is convenient in situations where we do not want or cannot impose %
a {\texttt{string-\-append}} call in staged code.\par

\subsection{Extensions}
\label{section4625}
\begin{figure}[ht]
\begin{scriptsize}
\setlength{\oldpretabcolsep}{\tabcolsep}
\addtolength{\tabcolsep}{-\tabcolsep}
{\setbox1 \vbox \bgroup
{\noindent \texttt{\begin{tabular}{l}
({\textcolor[rgb]{0.4117647058823529,0.34901960784313724,0.8117647058823529}{{\textbf{with-imported-modules}}}}\ (source-module-closure\\
\ \ \ \ \ \ \ \ \ \ \ \ \ \ \ \ \ \ \ \ \ \ \ \ '((guix\ build\ utils)))\\
\ \ \#$_{\mbox{\char126}}$(\textbf{\textrm{begin}}\\
\ \ \ \ \ \ {\textcolor[rgb]{0.4666666666666667,0.4,0.0}{{\textit{\textrm{;;\ Import\ the\ module\ in\ scope.}}}}}\\
\ \ \ \ \ \ ({\textcolor[rgb]{0.4117647058823529,0.34901960784313724,0.8117647058823529}{{\textbf{use-modules}}}}\ (guix\ build\ utils))\\
\\
\ \ \ \ \ \ {\textcolor[rgb]{0.4666666666666667,0.4,0.0}{{\textit{\textrm{;;\ Use\ a\ function\ from\ (guix\ build\ utils).}}}}}\\
\ \ \ \ \ \ (mkdir-p\ \#\$output)))\\
\end{tabular}
}}
\egroup{\box1}}%
\setlength{\tabcolsep}{\oldpretabcolsep}

\end{scriptsize}
\caption{\label{fig-gexp-modules}Specifying imported modules in a gexp.}\end{figure}
{\textbf{Modules.}} One of the reasons for using the same %
language uniformly is the ability to reuse Guile modules in %
several contexts.  Since builds are performed in an isolated %
environment, Scheme modules that are needed must be explicitly {\em{imported}} in that environment; in other words, the modules must be %
added as inputs to the derivation that needs them.  To that end, gexp %
objects embed information about the modules they need; the {\texttt{with-\-imported-\-modules}} syntactic form allows users to specify %
modules to import in the gexps that appear in its body.  The example %
in Figure \ref{fig-gexp-modules} creates a gexp that %
requires the {\texttt{(guix build utils)}} module and the modules it %
depends on in its execution environment.  The source of these modules %
is taken from the user’s search path and added to the store when {\texttt{gexp-\-$>$derivation}} is called.\par
Note that, to actually bring the module in scope, we %
still need to use Guile’s {\texttt{use-\-modules}} form.  We choose to not %
conflate the notion of modules-in-scope and that of imported-modules %
because some of the modules come with Guile itself; importing those %
from the user’s environment would make derivations sensitive to the %
version of Guile used on the “host side”.\par
{\textbf{Cross-compilation.}} Guix supports %
cross-compilation to other hardware architectures or operating %
systems, where packages are built by a {\em{cross-compiler}} that %
generates code for the target platform.  To support this, we must be %
able to distinguish between {\em{native}} dependencies—dependencies %
that run on the build system—and {\em{target}} %
dependencies—dependencies that can only run on the target system.  At %
the level of package definitions, the distinction is made by having %
two fields{\char58} {\texttt{inputs}} and {\texttt{native-\-inputs}}.  To allow gexps %
to express this distinction, we introduced a new “unquote” form, {\texttt{ungexp-\-native}}, abbreviated as {\texttt{\#+}}.\par
In a native build context, {\texttt{\#+}} is strictly %
equivalent to {\texttt{\#\$}}.  However, when cross-compiling, elements %
introduced with {\texttt{\#+}} are lowered to their {\em{native}} %
derivation, while elements introduced with {\texttt{\#\$}} are lowered to a %
derivation that {\em{cross-compiles}} them for the target.  To %
illustrate this, consider the gexp used to convert the bootloader’s %
background image to a suitable format, which resembles that of Figure %
\ref{fig-build-gexp}.  In a cross-compilation context, the %
expression that converts the image should use the {\em{native}} %
ImageMagick, not the target ImageMagick, which it would not be able to %
run anyway.  Thus, we write {\texttt{\#+imagemagick}} rather than {\texttt{\#\$imagemagick}}.  “Nativeness” propagates to all the values beneath %
{\texttt{\#+}}.\par

\section{Experience}
\label{experience}
Guix and GuixSD are used in production by individuals and %
organizations to deploy software on laptops, servers, and clusters. %
Deploying GuixSD involves staging hundreds of gexps. %
Introducing a new core mechanism in such a project can be both fruitful %
and challenging.  This section reports on our experience using gexps in %
Guix.\par

\subsection{Package Build Procedures}
\label{section4638}
As explained earlier, gexps appeared quite recently in %
the history of Guix.  Package definitions like that of Figure \ref{fig-package-def} rely on the previous ad-hoc staging %
mechanism, as can be seen in the use of labels in the {\texttt{inputs}} field of definitions.  Guix today includes more than 6,000 %
packages, which still use this old, pre-gexp style.  We are %
considering a migration to a new style but given the size of the %
repository, this is a challenging task and we must make sure every use %
case is correctly addressed in the new model.\par
In theory, labels are no longer needed with %
gexps since one can now use a {\texttt{\#\$}} escape to %
refer to the absolute file name of an input in {\texttt{arguments}}.  The indirection that %
labels introduced had one benefit though{\char58} one could create a package %
variant with a different {\texttt{inputs}} field, and {\texttt{(assoc-\-ref \%build-\-inputs …)}} %
calls in build-side code would automatically resolve to the new %
dependencies.  If we instead allow for direct use of {\texttt{\#\$}} in package %
{\texttt{arguments}}, those will be unaffected by changes in {\texttt{inputs}}.  It remains to %
be seen how we can allow {\texttt{\#\$}} forms while not sacrificing this %
flexibility.\par

\subsection{System Services}
\label{section4677}
GuixSD, the Guix-based GNU/Linux distribution, was one of %
the main motivations behind gexps.  The principle behind GuixSD is %
that, given a high-level {\texttt{operating-\-system}} declaration that %
specifies user accounts, system services, and other settings, the %
complete system is instantiated to the store.  To achieve that, a lot %
of files must be generated{\char58} start/stop script for all the system %
services (daemons, operations such as file system mounts, etc.), %
configuration files, and an initial RAM disk (or {\em{initrd}}) for %
the kernel Linux.\par
\begin{figure}[ht]
\begin{scriptsize}
\setlength{\oldpretabcolsep}{\tabcolsep}
\addtolength{\tabcolsep}{-\tabcolsep}
{\setbox1 \vbox \bgroup
{\noindent \texttt{\begin{tabular}{l}
(expression-$>$initrd\\
\ ({\textcolor[rgb]{0.4117647058823529,0.34901960784313724,0.8117647058823529}{{\textbf{with-imported-modules}}}}\ (source-module-closure\\
\ \ \ \ \ \ \ \ \ \ \ \ \ \ \ \ \ \ \ \ \ \ \ \ \ '((gnu\ build\ linux-boot)))\\
\ \ \ \#$_{\mbox{\char126}}$(\textbf{\textrm{begin}}\\
\ \ \ \ \ \ \ ({\textcolor[rgb]{0.4117647058823529,0.34901960784313724,0.8117647058823529}{{\textbf{use-modules}}}}\ (gnu\ build\ linux-boot))\\
\\
\ \ \ \ \ \ \ (boot-system\ \#{\char58}mounts\ '\#\$file-systems\\
\ \ \ \ \ \ \ \ \ \ \ \ \ \ \ \ \ \ \ \ \#{\char58}linux-modules\ '\#\$linux-modules\\
\ \ \ \ \ \ \ \ \ \ \ \ \ \ \ \ \ \ \ \ \#{\char58}linux-module-directory\ '\#\$kodir)))\\
\end{tabular}
}}
\egroup{\box1}}%
\setlength{\tabcolsep}{\oldpretabcolsep}

\end{scriptsize}
\caption{\label{fig-initrd}Creating a derivation that builds an initrd.}\end{figure}
The initrd is a small file system image that the kernel %
Linux mounts as its initial root file system.  It then runs the {\texttt{/\-init}} program therein; this program is responsible for mounting the %
real root file system and for loading any drivers needed to achieve %
that.  If the file system is encrypted, this is also the place where a %
“mapped device” is set up to decrypt it.  On GuixSD, this init program %
is a Scheme program that we generate based on the OS configuration, %
using gexps.  Figure \ref{fig-initrd} illustrates the %
creation of an initrd.  Here {\texttt{expression-\-$>$initrd}} returns a %
derivation that builds an initrd containing the given gexp as the {\texttt{/\-init}} program.  The staged program in this example calls the {\texttt{boot-\-system}} function from the {\texttt{(gnu build linux-\-boot)}} %
module.  The initrd is automatically populated with Guile and its %
dependencies, the closure of the {\texttt{(gnu build linux-\-boot)}} %
module, and the {\texttt{kodir}} store item which contains kernel modules %
(drivers).  That all the relevant store items referred to by the gexp, %
directly or indirectly, are “pulled” in the initrd comes for free.\par
Once the root file system is mounted, the initrd passes %
control to the Shepherd, our daemon-managing daemon\footnote{{\texttt{https{\char58}/\-/\-gnu.\-org/\-software/\-shepherd/\-}}}.  The Shepherd is responsible for %
starting system services—from the email or SSH daemon to the X %
graphical display server—and for doing other initialization operations %
such as mounting additional file systems.  The Shepherd is written in %
Scheme; its “configuration file” is a small Scheme program that %
instantiates service objects, each of which has a {\texttt{start}} and a %
{\texttt{stop}} method.  In GuixSD, service definitions take the form of %
a host-side structure with a {\texttt{start}} and {\texttt{stop}} field, %
both of which are gexps.  Those gexps are eventually {\em{spliced}} %
into the Shepherd configuration file.\par
Since this is all Scheme, and since Guix has a {\texttt{(gnu %
build linux-\-container)}} module to create Linux {\em{containers}} %
(isolated execution environments), we were able to reuse this %
container module within the Shepherd [4].  The only thing we had %
to do to achieve this was to (1) wrap our {\texttt{start}} gexp in {\texttt{with-\-imported-\-modules}} so that it has access to the container %
functionality, and (2) use our start-process-in-container function %
lieu of the Shepherd’s own start-process function.  This is a good %
example of cross-stage code sharing, where the second stage in this %
case is the operating system’s run-time environment.\par

\subsection{System Tests}
\label{section4710}
\begin{figure}[ht]
\begin{scriptsize}
\setlength{\oldpretabcolsep}{\tabcolsep}
\addtolength{\tabcolsep}{-\tabcolsep}
{\setbox1 \vbox \bgroup
{\noindent \texttt{\begin{tabular}{l}
\#$_{\mbox{\char126}}$(\textbf{\textrm{begin}}\\
\ \ \ \ ({\textcolor[rgb]{0.4117647058823529,0.34901960784313724,0.8117647058823529}{{\textbf{use-modules}}}}\ (gnu\ build\ marionette)\\
\ \ \ \ \ \ \ \ \ \ \ \ \ \ \ \ \ (srfi\ srfi-64)\ (ice-9\ \textbf{\textrm{match}}))\\
\\
\ \ \ \ {\textcolor[rgb]{0.4666666666666667,0.4,0.0}{{\textit{\textrm{;;\ Spawn\ the\ VM\ that\ runs\ the\ declared\ OS.}}}}}\\
\ \ \ \ ({\textcolor[rgb]{0.4117647058823529,0.34901960784313724,0.8117647058823529}{{\textbf{define}}}}\ marionette\ (make-marionette\ (list\ \#\$vm)))\\
\\
\ \ \ \ (test-begin\ {\textcolor[rgb]{1.0,0.0,0.0}{"basic"}})\\
\ \ \ \ (test-assert\ {\textcolor[rgb]{1.0,0.0,0.0}{"uname"}}\\
\ \ \ \ \ \ (\textbf{\textrm{match}}\ (marionette-eval\ '(uname)\ marionette)\\
\ \ \ \ \ \ \ \ (\#({\textcolor[rgb]{1.0,0.0,0.0}{"Linux"}}\ host-name\ version\ \_\ architecture)\\
\ \ \ \ \ \ \ \ \ (and\ (string=?\ host-name\\
\ \ \ \ \ \ \ \ \ \ \ \ \ \ \ \ \ \ \ \ \ \ \ \ \#\$(operating-system-host-name\ os))\\
\ \ \ \ \ \ \ \ \ \ \ \ \ \ (string-prefix?\ \#\$(package-version\\
\ \ \ \ \ \ \ \ \ \ \ \ \ \ \ \ \ \ \ \ \ \ \ \ \ \ \ \ \ \ \ \ \ (operating-system-kernel\ os))\\
\ \ \ \ \ \ \ \ \ \ \ \ \ \ \ \ \ \ \ \ \ \ \ \ \ \ \ \ \ \ version)\\
\ \ \ \ \ \ \ \ \ \ \ \ \ \ (string-prefix?\ architecture\ \%host-type)))))\\
\ \ \ \ (test-end)\\
\ \ \ \ (exit\ (=\ (test-runner-fail-count\ (test-runner-current))\ 0))))\\
\end{tabular}
}}
\egroup{\box1}}%
\setlength{\tabcolsep}{\oldpretabcolsep}

\end{scriptsize}
\caption{\label{fig-system-test}Core of a whole-system test.}\end{figure}
GuixSD comes with a set of {\em{whole-system tests}}. %
Each of them takes an {\texttt{operating-\-system}} definition, which defines %
the OS configuration, instantiates it in a virtual machine (VM), and %
verifies that the system running in the VM matches some of the settings.  The %
guest OS is instrumented with a Scheme interpreter that evaluates %
expressions sent by the host OS—we call it “marionette”.\par
Whole-system tests are derivations whose build programs are %
gexps like that of Figure \ref{fig-system-test}. %
The build program passes {\texttt{vm}}, the script to spawn the VM, to the %
instrumentation tool.  The test then uses {\texttt{marionette-\-eval}} to %
call the {\texttt{uname}} function in the guest{\char58} an {\em{additional code stage}} is %
introduced here, this time using {\texttt{quote}} since gexps are currently %
limited to contexts with a connection to the build daemon.  The test matches the %
return value of {\texttt{uname}} against the expected vector, and makes %
sure the information corresponds to the various bits declared in {\texttt{os}}, our OS definition.\par

\section{Limitations}
\label{limitations}
{\textbf{Hygiene.}} Our implementation of hygiene, discussed %
in Section~3.2, follows the %
well-documented approach to the problem [13, 15].  Rhiger’s %
implementation handles a single binding construct ({\texttt{lambda}}) and %
MetaScheme handles a couple more constructs, but ours has %
to deal with more binding constructs{\char58} R6RS defines around ten %
binding constructs, and Guile adds a couple more.\par
Hygiene in multi-stage programs relies on identifying binding %
constructs.  This turns out to be hard to achieve in Scheme because %
macros can define {\em{new}} bindings constructs. %
Our \begin{math}\alpha\end{math}-renaming pass is oblivious to those so it will %
not properly rename bindings introduced by user-defined macros.  The %
macro expander, of course, does this and more already, so it would be %
tempting to reuse it rather than duplicate part of its work.  However, %
we do not want to macro-expand staged code; instead, macro expansion %
should be performed “the normal way”, by the Guile program that %
compiles or evaluates the staged code.  Again, this ensures %
reproducibility across Guix installations since we control precisely %
the Guile variant used in derivations whereas we do not control the %
Guile variant used to evaluate “host-side” code.  How we could hook %
into Guile’s macro expander, based on {\texttt{psyntax}} [7], is still an open question.  To our %
knowledge, this problem of hygienic staging of a language with macros %
has not been addressed in literature outside of work on macro expanders %
[7].\par
On top of that, {\texttt{gexp}} must track the {\em{quotation level}} of several types of quotation{\char58} {\texttt{gexp}}, {\texttt{quote}}, {\texttt{quasiquote}}, and {\texttt{syntax}} (though our %
implementation currently leaves out {\texttt{syntax}} handling).  For %
each quotation type, \begin{math}\alpha\end{math}-renaming must be skipped when %
the quotation level is greater than zero.  For example, in {\texttt{\#$_{\mbox{\char126}}$(lambda (x) `(x ,x))}}, the first {\texttt{x}} must {\em{not}} be %
renamed, while the second one must be renamed.  Needless to say, the %
resulting implementation lacks the conciseness of those found in the %
literature.  This is another area that could use help from the macro %
expander.\par
{\textbf{Modules in scope.}}  The {\texttt{with-\-imported-\-modules}} form allows to specify which modules a gexp %
expects in its execution environment, but we currently lack a way to %
specify {\em{which modules should be in scope}}, which could be %
useful in some situations.  Part of the reason is that in Guile {\texttt{use-\-modules}} clauses must appear at the top level, and thus they %
cannot be used in a gexp that ends up being inserted in a %
non-top-level position.  Macro expanders know the modules in scope at %
macro-definition points so they can replace free variables in residual %
code with fully-qualified references to variables inside the modules %
in scope at the macro definition point.  How to achieve something %
similar with gexp, which lack the big picture that a macro expander has, %
remains an open question.\par
{\textbf{Cross-stage debugging.}} {\texttt{gexp-\-$>$derivation}} %
emits build programs as sexps in a file in {\texttt{/\-gnu/\-store}}.  When %
an error occurs during the execution of these programs, Guile prints a %
backtrace that refers to source code locations {\em{inside the %
generated code}}.  What we would like, instead, is for the backtrace %
to refer to the location {\em{of the gexp itself}}.  C has {\texttt{\#line}} directives, which code generators insert in generated code to %
{\em{map}} generated code to its source.  If a similar %
feature was available in Scheme, it would be unsuitable{\char58} moving the %
source code where a gexp appears would lead to a different derivation, %
in turn triggering a rebuild of everything that depends on it. %
Instead we would need a way to pass source code %
mapping information {\em{out-of-band}}, in a way that does not affect %
the derivation that is produced.  We are investigating ways to %
achieve that.\par

\section{Related Work}
\label{related}
Like Guix, Nix must be %
able to include references to store items (derivation results) in %
generated code while keeping track of derivations this generated %
code depends on.  However, Nix is a single-stage language{\char58} %
the “build side” is left to other languages such as %
Bash or Perl.  Users can splice Nix expressions in %
strings using {\em{string interpolation}} %
[6]; the interpreter records this dependency in the string %
context and substitutes the reference with the output file name of the %
derivation.\par
Nix views staged code as mere strings and thus %
does not provide any guarantee on the generated code. %
The string interpolation syntax ({\texttt{\$\{}}…{\texttt{\}}} %
sequences) often clashes with the target’s language syntax (e.g., %
Bash uses dollar-brace syntax to reference variables), which can lead %
to subtle errors.  In addition, comments and whitespace in those  %
strings are preserved, and %
changing those triggers a rebuild of the derivation, which is %
inconvenient.\par
GuixSD and MirageOS both %
aim to unify configuration and deployment into a single high-level %
language framework [14].  MirageOS %
uses code staging through MetaOCaml, though that is limited to the %
implementation of its data storage layer.\par
Code staging is often studied in the context of optimized code %
generation [1, 10, 11], %
or that of hygienic macros [7, 8, 9].  Gexps appear to be the first %
use of staging in the context of software deployment.  Apart from LMS, %
which relies on types [11], most approaches to %
staging rely on syntactic annotations similar to {\texttt{bracket}} or {\texttt{gexp}}.  Scheme’s {\em{hygienic macros}}, now part of the R5RS and %
R6RS standards, as well as MacroML [8] support %
user-defined binding constructs; the macro expander recognizes those %
bindings constructs, which allows it to track bindings and preserve %
hygiene, notably by \begin{math}\alpha\end{math}-renaming introduced bindings.\par
MetaScheme is a translation of MetaOCaml’s staging %
primitives, {\texttt{bracket}}, {\texttt{escape}}, and {\texttt{lift}} [15] implemented as a macro that expands to an %
sexp.  It considers only a few %
core binding constructs and does not address hygiene in the presence %
of user-defined binding constructs introduced by macros. %
Rhiger’s work [13] follows a similar %
approach but redefines Scheme’s quasiquotation instead of %
introducing new constructs.\par
Staged Scheme, or S\begin{math}\sp{\mbox{2}}\end{math}, provides bracket, escape, and lift forms separate %
from {\texttt{quasiquote}} and {\texttt{unquote}} [10]. %
As with {\texttt{syntax-\-case}} [7] and {\texttt{gexp}}, staged code has a disjoint type, as opposed to being a list. %
S\begin{math}\sp{\mbox{2}}\end{math}’s focus is on programs with %
possibly more than two stages, whereas gexps are, in practice, used for %
two-stage programs.  The article discusses {\em{code regeneration}} %
at run time; gexps have a similar requirement here{\char58} at run time a %
given gexp may be instantiated for different systems, for %
instance {\texttt{x86\_64-\-linux}} and {\texttt{i686-\-linux}}.\par
Hop performs {\em{heterogenous staging}}{\char58} the source language is Scheme, but the %
generated code is JavaScript [12].  In Hop %
{\texttt{$_{\mbox{\char126}}$}} introduces staged client-side expressions and %
{\texttt{\$}} escapes to unstaged server-side code.  Unlike %
gexps, support for {\texttt{$_{\mbox{\char126}}$}} forms is built in the Hop compiler, %
and {\texttt{$_{\mbox{\char126}}$}} forms are not first-class objects.  Hop comes with useful %
multi-stage debugging facilities not found in Guix, such as the %
ability to display cross-stage stack traces with correct source %
location information.  It also has a way to express modules in scope for %
staged code.\par

\section{Conclusion}
\label{conclusion}
G-expressions are a novel application of hygienic code %
staging techniques from the literature to functional software %
deployment.  They extend common staging constructs ({\texttt{bracket}} %
and {\texttt{escape}}) with additional tooling{\char58} cross-compilation-aware %
escapes, and imported-module annotations.  Gexps are used in %
production to express package build procedures in Guix as well as all %
the assembly of operating system components in GuixSD.  Using a %
single-language framework with staging has proved to enable new ways %
of code reuse and composition.\par

\newpage
\balance
 %
 %

\section{References}
\label{chapter4807}
\begin{flushleft}{%
\sloppy
\sfcode`\.=1000\relax
\newdimen\bibindent
\bibindent=0em
\begin{list}{}{%
        \settowidth\labelwidth{[21]}%
        \leftmargin\labelwidth
        \advance\leftmargin\labelsep
        \advance\leftmargin\bibindent
        \itemindent -\bibindent
        \listparindent \itemindent
        \itemsep 0pt
    }%
\item[{\char91}1{\char93}] Baris Aktemur, Yukiyoshi Kameyama, Oleg Kiselyov, and Chung-chieh Shan. Shonan Challenge for Generative Programming{\char58} Short Position Paper. In {\textit{Proceedings of the ACM SIGPLAN 2013 Workshop on Partial Evaluation and Program Manipulation}}, PEPM '13, pp. 147--154, ACM, 2013.
\item[{\char91}2{\char93}] Alan Bawden. Quasiquotation in Lisp. In {\textit{Proceedings of the ACM SIGPLAN Workshop on Partial %
Evaluation and Semantics-Based Program Manipulation (PEPM 1999)}}, pp. 4--12, 1999.
\item[{\char91}3{\char93}] Ludovic Court\`{e}s. Functional Package Management with Guix. In {\textit{European Lisp Symposium}}, June 2013.
\item[{\char91}4{\char93}] Ludovic Court\`{e}s. Running system services in containers. April 2017. {\textit{https{\char58}//gnu.org/s/guix/news/running-system-services-in-containers.html}}.
\item[{\char91}5{\char93}] Eelco Dolstra, Merijn de Jonge, Eelco Visser. Nix{\char58} A Safe and Policy-Free System for Software Deployment. In {\textit{Proceedings of the 18th Large Installation System Administration Conference (LISA '04)}}, pp. 79--92, USENIX, November 2004.
\item[{\char91}6{\char93}] Eelco Dolstra, Andres L\"{o}h, Nicolas Pierron. NixOS{\char58} A Purely Functional Linux Distribution. In {\textit{Journal of Functional Programming}}, (5-6) , New York, NY, USA, November 2010, pp. 577--615.
\item[{\char91}7{\char93}] R. Kent Dybvig. Writing Hygienic Macros in Scheme with Syntax-Case. Technical Report 356, Indiana Computer Science Department, June 1992.
\item[{\char91}8{\char93}] Steven E. Ganz, Amr Sabry, and Walid Taha. Macros As Multi-stage Computations{\char58} Type-safe, Generative, Binding Macros in MacroML. In {\textit{Proceedings of the Sixth ACM SIGPLAN International Conference on Functional Programming}}, ICFP '01, pp. 74--85, ACM, 2001.
\item[{\char91}9{\char93}] Eugene Kohlbecker, Daniel P. Friedman, Matthias Felleisen, and Bruce Duba. Hygienic Macro Expansion. In {\textit{Proceedings of the 1986 ACM Conference on LISP and Functional Programming}}, LFP '86, pp. 151--161, ACM, 1986.
\item[{\char91}10{\char93}] Zhenghao Wang and Richard R. Muntz. Managing Dynamic Changes in Multi-stage Program Generation Systems. In {\textit{Proceedings of Generative Programming and Component Engineering (GPCE)}}, Springer Berlin Heidelberg, Batory, Don and Consel, Charles and Taha, Walid (editor), pp. 316--334, October 2002.
\item[{\char91}11{\char93}] Tiark Rompf and Martin Odersky. Lightweight Modular Staging{\char58} A Pragmatic Approach to Runtime Code Generation and Compiled DSLs. In {\textit{Commun. ACM}}, 55(6) , New York, NY, USA, June 2012, pp. 121--130.
\item[{\char91}12{\char93}] Manuel Serrano and Christian Queinnec. A Multi-tier Semantics for Hop. In {\textit{Higher Order Symbol. Comput.}}, 23(4) , Hingham, MA, USA, November 2010, pp. 409--431.
\item[{\char91}13{\char93}] Morten Rhiger. Hygienic Quasiquotation in Scheme. In {\textit{Proceedings of the 2012 Annual Workshop on Scheme and Functional Programming}}, Scheme '12, pp. 58--64, ACM, 2012.
\item[{\char91}14{\char93}] Anil Madhavapeddy and David J. Scott. Unikernels{\char58} Rise of the Virtual Library Operating System. In {\textit{Queue}}, 11(11) , New York, NY, USA, December 2013, pp. 30{\char58}30--30{\char58}44.
\item[{\char91}15{\char93}] Oleg Kiselyov and Chung-chieh Shan. MetaScheme, or untyped MetaOCaml. August 2008. {\textit{http{\char58}//okmij.org/ftp/meta-programming/\#meta-scheme}}.

\end{list}}
\end{flushleft}

\end{document}

%% file: categories.tex
